\documentclass[aps,prd,twocolumn,superscriptaddress]{revtex4}

\usepackage{hyperref}
\usepackage{amsmath,amssymb,amsfonts}
\usepackage{color}
\usepackage{graphicx}
\usepackage{enumerate} 
\usepackage{colordvi} 
\usepackage{bm}

\newcommand{\gpc}{\, {\rm Gpc}}
\newcommand{\mpc}{\, {\rm Mpc}}

\newcommand{\hgpc}{\, h^{-1} \gpc}
\newcommand{\hmpc}{\, h^{-1} \mpc}

\newcommand{\vx}{\mathbf{x}}

\newcommand{\vecr}{\mathbf{r}}

\newcommand{\vv}{\mathbf{v}}

\newcommand{\wt}{\widetilde}

\newcommand{\be}{\begin{equation}}
\newcommand{\ee}{\end{equation}}
\newcommand{\bey}{\begin{eqnarray}}
\newcommand{\eey}{\end{eqnarray}}

\newcommand{\nn}  {\nonumber}

\begin{document}
\title{
Splashback radius of nonspherical dark matter halos \\
from cosmic density and velocity fields
}
\author{Teppei Okumura}\email{tokumura@asiaa.sinica.edu.tw}
\affiliation{Institute of Astronomy and Astrophysics, Academia Sinica, P. O. Box 23-141, Taipei 10617, Taiwan}
\affiliation{Kavli Institute for the Physics and Mathematics of the Universe (WPI), UTIAS, The University of Tokyo, Kashiwa, Chiba 277-8583, Japan}

\author{Takahiro Nishimichi}
\affiliation{Kavli Institute for the Physics and Mathematics of the Universe (WPI), UTIAS, The University of Tokyo, Kashiwa, Chiba 277-8583, Japan}
\affiliation{CREST, JST, 4-1-8 Honcho, Kawaguchi, Saitama, 332-0012, Japan}

\author{Keiichi Umetsu}
\affiliation{Institute of Astronomy and Astrophysics, Academia Sinica, P. O. Box 23-141, Taipei 10617, Taiwan}

\author{Ken Osato}
\affiliation{Department of Physics, University of Tokyo, 7-3-1 Hongo, Bunkyo-ku, Tokyo 113-0033 Japan}

\date{\today}

\begin{abstract}
We investigate the splashback features of dark-matter halos based on
cosmic density and velocity fields. Besides the density correlation
function binned by the halo orientation angle, which was used in the
literature, we introduce, for the first time, the corresponding
velocity statistic, alignment momentum correlation function, to take into account the asphericity of halos. 
Using large-volume, high-resolution $N$-body simulations, we measure the
alignment statistics of density and velocity. On halo scales, $x\sim
R_\mathrm{200m} \sim 1\hmpc$, we detect a sharp steepening in the
momentum correlation associated with the physical halo boundary, or
the splashback feature, which is found more prominent than in the
density correlation.  We also find that the splashback radius
determined from the density correlation becomes $\sim 3.5\%$ smaller
than that from the momentum correlation, with their correlation coefficient
being 0.605.  Moreover, the orientation-dependent splashback feature
due to halo asphericity is measured when the density profile is
determined by dark-matter particles, which can be used as a test of
collisional cold dark matter since the halo shape is predicted to be rounder in such a model. 
\end{abstract}
\maketitle

\flushbottom
\section{Introduction}\label{sec:intro}
In the current paradigm of cosmic structure formation, galaxies, which
are observed as a tracer of the large-scale structure of the Universe,
are considered to be formed within dark-matter halos. Besides,
modeling a halo power spectrum is the important first step to properly
interpreting the observed galaxy clustering, from which to extract
cosmological information.  Dark-matter halos thus play a fundamental
role in both structure formation and cosmological studies (e.g., Refs.
\cite{Cooray:2002,Mo:2010}).

Recently, the phase-space structure in halo outskirts has been
extensively studied based on $N$-body simulations, leading to the
discovery of a steepening in the outer density profile of dark-matter
halos \cite{Diemer:2014}.  This feature is interpreted as a sharp
density enhancement associated with the orbital apocenter of the
recently accreted matter in the growing halo potential. The location
of this steepening is referred to as the splashback radius,
$R_\mathrm{sp}$, and depends on cosmology as well as on halo mass and
redshift.  The splashback radius provides a physical boundary of halos
\cite{Diemer:2014,Adhikari:2014,More:2015a,Shi:2016}, and is related
to the transition scale between the one-halo and two-halo regimes in the
galaxy power spectrum or correlation function to a certain extent
\cite{Cooray:2002,More:2015a}.

Using efficient cluster-finding algorithms based on the observed galaxy
distribution \cite{Rykoff:2014,Oguri:2014}, the splashback features have
been studied by observing the galaxy density profile
and weak lensing profile
\cite{More:2016,Umetsu:2017,Baxter:2017,Chang:2017} (see Ref.
\cite{Busch:2017} for difficulties in observing the splashback
feature).  Further studies revealed that dynamical friction acting on
massive subhalos orbiting in their parent clusters makes splashback
features appear at smaller cluster radii
\cite{Adhikari:2016,Diemer:2017,Diemer:2017a}.  However, the detected
splashback radius is found to be significantly smaller than predicted
by $N$-body simulations, even though the effect of dynamical friction
is considered \cite{More:2016,Chang:2017}. Thus, careful work is
required both from theoretical and observational aspects.

Splashback features are determined by the orbits of dark matter around
halo centers and thus fully characterized in phase space.  Hence, the
commonly used density statistic alone cannot capture the full
dynamical information.  Furthermore, the two-halo term of the density
statistic is enhanced by the galaxy bias,
whereas its impact on the
determination of the splashback radius has not been discussed in the
literature.  Another important fact on precisely measuring the
splashback radius is that halos are aspherical.  Thus, spherical
averaging would smear out the splashback features
\cite{Mansfield:2017}.  While the caustic techniques have been
extensively studied in phase space to measure dynamical mass profiles
of clusters from infall velocity patterns
\cite{Diemand:2008,Lemze:2009,Vogelsberger:2011,Rines:2013,Svensmark:2015},
these analyses have not been performed in the context of splashback
studies. In this paper, we present a detailed study of splashback
features based on both density and velocity statistics, focusing on
the issues described above.

This paper is organized as follows. 
In Sec. \ref{sec:theory}, we present the formalism of alignment density and velocity statistics used to study the splashback features.
Section \ref{sec:nbody} describes the $N$-body simulations and how we construct mock cluster and galaxy samples. 
Section \ref{sec:analysis} presents measurements of alignment density and momentum correlation functions, their splashback features, and constraints on the splashback radius. 
Our conclusions are given in Sec. \ref{sec:conclusions}.

\section{Formalism}\label{sec:theory}
The three-dimensional density profile around clusters is computed by the cross-correlation functions between halo centers and mass tracers. 
When dark matter particles and galaxies are used as the tracers, 
the cross-correlations are, respectively, expressed as
$\xi_{mc}(r)=\left\langle \delta_m(\vx_1)\delta_c(\vx_2)\right\rangle$ and
$\xi_{gc}(r)=\left\langle \delta_g(\vx_1)\delta_c(\vx_2)\right\rangle$ (e.g., see Ref. \cite{Hayashi:2008}), where
$r=|{\bf r}|=|\vx_2-\vx_1|$ and $\delta_c$, $\delta_g$, and $\delta_m$ are the overdensity fields traced by clusters, galaxies, and matter, respectively.
In weak lensing and galaxy redshift surveys, one can, respectively, observe 
the weak lensing profile, $\Sigma(R)$ [or $\Delta\Sigma(R)$], and 
the galaxy density profile, $\Sigma_g(R)$, which are the line-of-sight projection of the cross-correlation functions, $\xi_{mc}({\bf r})$ and $\xi_{gc}({\bf r})$.

\subsection{Alignment density correlation}
To take into account the asphericity of dark matter halos, we consider the angle-binned or alignment correlation function \cite{Paz:2008,Faltenbacher:2009,Osato:2018}, an extension of the conventional matter and galaxy density profiles around clusters, respectively, $\xi_{mc}(r)$ and $\xi_{gc}(r)$,
by taking account of the orientations of the clusters,
\be
\xi_{Ac}(r,\theta)=\left\langle \delta_A({\bf x}_1,\theta)\delta_c({\bf x}_2)\right\rangle, \label{eq:align_d}
\ee
where $A=\{m,g\}$. 
Here, $\theta$ is the angle between the elongated orientation of cluster
halos, defined by the major axis of ellipsoidal halo shapes,
and the separation vector ${\bf r}$.
The conventional correlation function 
can be obtained by integrating over $\theta$,
\be
\xi_{Ac}(r)=\int^{1}_{0} d\cos{\theta} \xi_{Ac}(r,\theta).
\ee 

The alignment correlation is related to the density-ellipticity correlation, 
a main source of contamination for measurements of the
gravitational shear power spectrum in weak lensing surveys, also known as
intrinsic alignments, 
$\xi_{g+}({\bf r})=\left\langle\delta_g(\vx_1)\left[1+\delta_c(\vx_2)\right]\gamma^I(\vx_2)\right\rangle$, 
where 
$\gamma^I(\vx)=\frac{1-q^2}{1+q^2}\cos(2\theta_{p})$, $\theta_{p}$ is the angle projected onto the celestial sphere, and $q$ is the minor-to-major-axis ratio of halos 
\cite{Catelan:2001,Hirata:2004,Mandelbaum:2006,Okumura:2009,Schafer:2009,Troxel:2015,Joachimi:2015,van-Uitert:2017}.
This function is related to the alignment
correlation function by
\be
\wt{\xi}_{g+}(\vecr)=(2/\pi)\int^{\pi/2}_{0}d\theta \cos(2\theta_{p})\xi_{gc}(\vecr,\theta_{p}),
\ee 
where $\wt{\xi}_{g
+}$ is the same as $\xi_{g +}$ but with $q$ fixed to $q=0$
\cite{Okumura:2009a,Faltenbacher:2009,Blazek:2011}. While these two
statistics are complimentary to each other, we will focus on the
alignment correlation function because it provides direct
insight on how the matter is distributed along and perpendicular to the
major axis of halos.  

\subsection{Alignment velocity statistics}
Next we consider a statistic
with respect to the cosmic velocity field, the momentum correlation
function
\cite{Gorski:1988,Fisher:1995,Strauss:1995,Okumura:2014,Sugiyama:2016},
$\psi_{Ac}(r)=\left\langle \left[1+\delta_A ({\bf
    x}_1)\right]\left[1+\delta_c ({\bf x}_2)\right] \vv_{A} ({\bf
  x}_1)\cdot \vv_{c} ({\bf x}_2)\right\rangle$,
  where $\vv_A$ is the peculiar velocity of field $A$ (namely, the cosmic expansion term is not included)
\footnote{To convert this to the one related to the observable
  quantity, one needs to replace the $(\vv_g\cdot\vv_c)$ term by
  $(v_g^zv_c^z)$, where $v_A^z(\vecr)$ is the radial component of the
  peculiar velocity of sample $A$, $v_A^z(\vecr)=\vv_A(\vecr)\cdot
  \hat{\vecr}$.  Then the result will simply differ by a factor of
  $3$.}.
We propose using this momentum correlation as a probe of the splashback radius
because the splashback features are fully characterized in phase space. 
In analogy to the density statistic, we define the alignment momentum
correlation, $\psi_{Ac}(r,\theta)$, by replacing $\delta_A(\vx_1)$
in the above equation by $\delta_A(\vx_1,\theta)$,
\bey
\psi_{Ac}(r,\theta)=\left\langle \left[1+\delta_A ({\bf
    x}_1,\theta)\right]\left[1+\delta_c ({\bf x}_2)\right] 
\right. \ \ \ \ \ \ \ \ \ \ \ \nn \\ \left. \times    
    \vv_{A} ({\bf
  x}_1)\cdot \vv_{c} ({\bf x}_2)\right\rangle. \label{eq:align_p}
\eey
This and Eq. (\ref{eq:align_d}) are the main statistics we use to investigate the splashback features of nonspherical dark halos. 
Similarly to the
density case, the conventional momentum correlation, $\psi_{Ac}(\vecr)$,
can be obtained by averaging Eq. (\ref{eq:align_p}) over $\theta$,
\be
\psi_{Ac}(r)=\int^{1}_{0} d\cos{\theta} \psi_{Ac}(r,\theta).
\ee

We also introduce the angle-binned, density-weighted pairwise velocity
dispersion:
\bey
\sigma_{v,Ac}^2(r,\theta)
=\left\langle \left[1+\delta_A ({\bf x}_1,\theta)\right]
\left[1+\delta_c ({\bf x}_2)\right] 
\right. \ \ \ \ \ \ \ \ \ \ \ \nn \\ \left. \times
\left| \vv_{A} ({\bf x}_1)-\vv_{c} ({\bf x}_2)\right|^2\right\rangle.
\eey
However, its behavior is found to be essentially similar to that of
$\psi_{Ac}$, and hence we do not present this statistic in this paper. 
Moreover, we define the angle-binned pairwise infall
momentum,
$p_{Ac}(r,\theta)=\left\langle \left[1+\delta_A ({\bf
x}_1,\theta)\right]\left[1+\delta_c ({\bf x}_2)\right]\left[\vv_{A}
({\bf x}_1)-\vv_{c} ({\bf x}_2)\right]\cdot \hat{\vecr}\right\rangle$,
where the hat denotes a unit vector. Since the splashback feature is
smeared out in this statistic by construction, we do not show it in
this paper
\footnote{While we comprehensively analyze these statistics in
  Ref. \cite{Okumura:2018}, some results have been presented in the early
  version of this paper. See Ref. \cite{Okumura:2017a}.
  }.

\begin{table*}[bt!]
\caption{Properties of mock central/satellite subhalo samples at
  $z=0.306$. $f_{\rm sat}$ is the number fraction of satellites,
  $M_\mathrm{min}$ and $\overline{M}$ are the minimum and average masses of central subhalos
 in units of $10^{12}h^{-1}M_\odot$, respectively; 
  $\overline{n}$ is the number density in units of $h^3{\rm
    Mpc}^{-3}$; and $b_A$ ($A=\{c,g\}$) is the cluster/galaxy bias computed at the large-scale limit.}
\begin{center}
\begin{tabular}{l|c|ccccc}
\hline\hline
Halo types & Label & $f_{\rm sat}$ & $M_{\rm min}$ &   $\overline{n}$ &$b_A$ & $\overline{M}$ \\ 
\noalign{\hrule height 1pt}
Clusters (central)     &$c$ & 0 &100 & $ 2.05 \times 10^{-5}$ & $3.05$ & 188\\
Galaxies (central + satellite) \ \ \  &$g$ & \ \ \ 0.137 \ \ \ & \ \ \ 0.224 \ \ \ &  \ \ \  $5.58\times 10^{-4}$ \ \ \ &\ \ \ $1.69$ \ \ \ & \ \ \ 27.9 \ \ \ \\ 
\hline\hline
\end{tabular}
\end{center}
\label{tab:halo}
\end{table*}

\section{$N$-body simulations}\label{sec:nbody}
To study splashback features,
we use a series of large and high-resolution $N$-body simulations of
the $\Lambda$CDM cosmology seeded with Gaussian initial
conditions. These are performed as a part of the \texttt{dark
  emulator} project \cite{Nishimichi:2018}. We adopt the cosmological
parameters of $\Omega_m=1-\Omega_\Lambda = 0.315$, $\Omega_b=0.0492$,
$h=0.673$, $n_s=0.965$, and $\sigma_8=0.8309$. We employ $n_p=2048^3$
particles of mass $m_p= 1.02\times 10^{10}M_\odot / h$ in a cubic box
of side $L_{\rm box} = 1\hgpc$. We use 24 realizations in total and
analyze the snapshots at $z=0.306$.

Subhalos are identified using the R{\footnotesize OCKSTAR}
algorithm \cite{Behroozi:2013} from phase-space information of matter
particles. 
The velocity of each subhalo is determined by the average
particle velocity within the innermost 10\% of the subhalo radius.
We use the standard definition for the halo radius and mass, 
\be 
M_h \equiv M_{\rm \Delta m} = M(<R_{\rm\Delta m}) = (4\pi/3)\Delta
\rho_{\rm m}(z) R_{\rm\Delta m}^3,
\ee
where $\rho_{\rm m}$ is the mean mass
density of the Universe at given redshift $z$, and we adopt $\Delta=
200$. 
In the R{\footnotesize OCKSTAR} algorithm, if one subhalo is within the virial radius of another subhalo but the opposite is not the case, 
the latter halo is labeled as a central subhalo. 
On the other hand, if two or more subhalos are located within the virial radius of each other, 
the most massive one is labeled as a central subhalo, and another is labeled as (a) satellite subhalo(s).

To study cluster-scale halos, we select central subhalos with
$M_{h} \geq 10^{14}h^{-1}M_\odot$, which roughly corresponds to the
typical threshold of the richness parameters used by the
cluster-finding algorithms in the literature.  We create mock galaxy
catalogs using a halo occupation distribution (HOD) model
\cite{Zheng:2005} applied for the LOWZ galaxy sample of the SDSS-III
Baryon Oscillation Spectroscopic Survey obtained by
Ref. \cite{Parejko:2013}. We populate halos with galaxies according to the
best-fitting HOD $N(M_h)$. For halos that contain satellite galaxies,
we randomly draw $N(M_h)-1$ member satellite subhalos to mimic the
positions and velocities of the satellites (see Refs. 
\cite{Nishimichi:2014,Okumura:2017} for alternative methods). 
We use a random selection of subhalos rather than the largest subhalos,
because a satellite subhalo undergoes the effect of a tidal disruption in the host halo
and its mass decreases as it goes toward the center of the gravitational potential.  
Thus, if we selected the largest satellite subhalos to host galaxies, we would 
preferentially pick up the satellite subhalos residing it the outskirts of the gravitational potential, which conflicts with the galaxy distribution in observation.
For this
HOD prescription we use all the subhalos with 20 or more than
particles in the simulation box. We assume central subhalos to have
triaxial shapes and estimate the orientations of their major axes
using the second moments of the mass distribution 
\cite{Jing:2002a}. Table \ref{tab:halo} summarizes
properties of our mock samples.

\begin{figure*}[bt]
\includegraphics[width=0.45\textwidth,angle=0,clip]{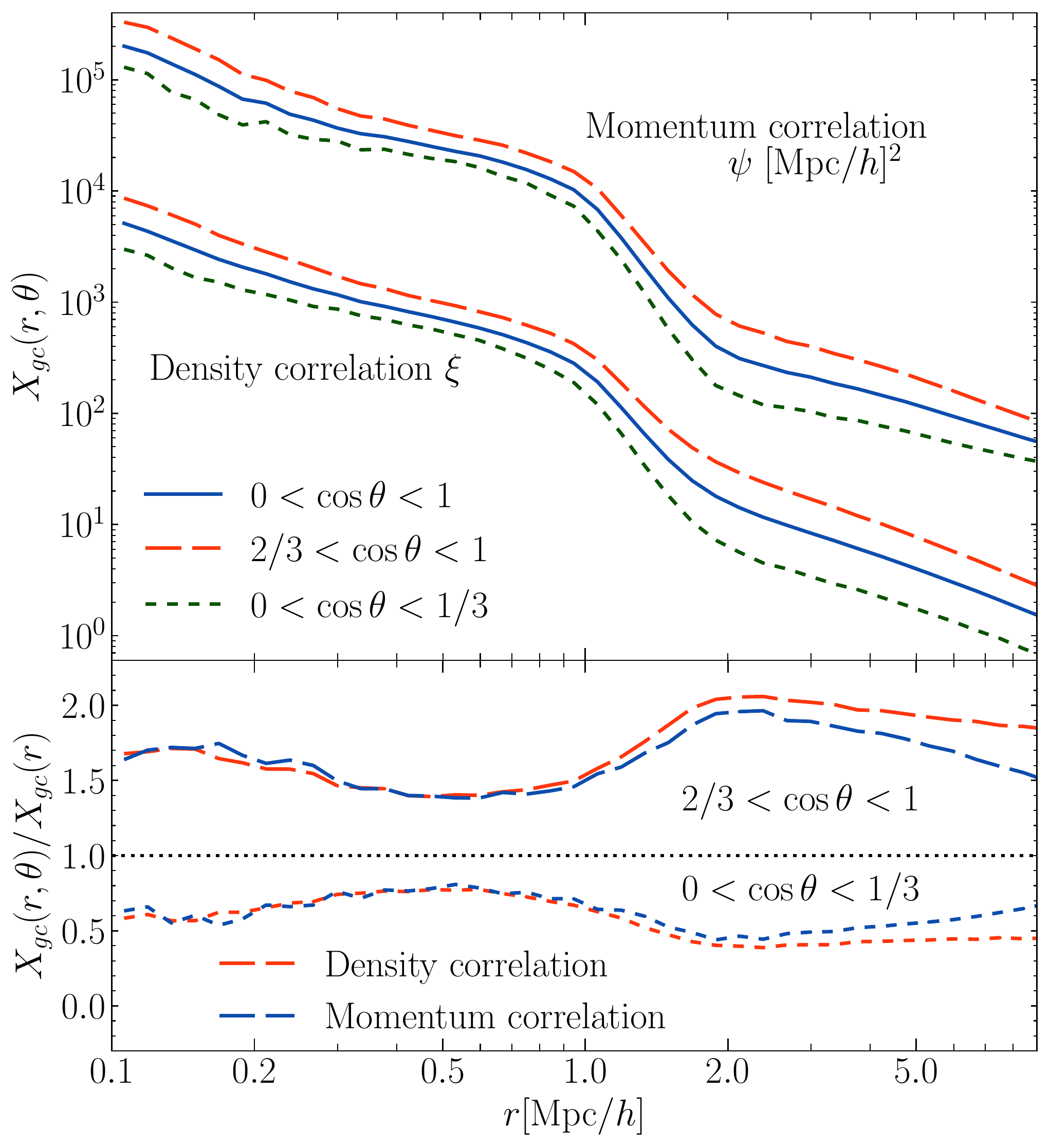}
\includegraphics[width=0.45\textwidth,angle=0,clip]{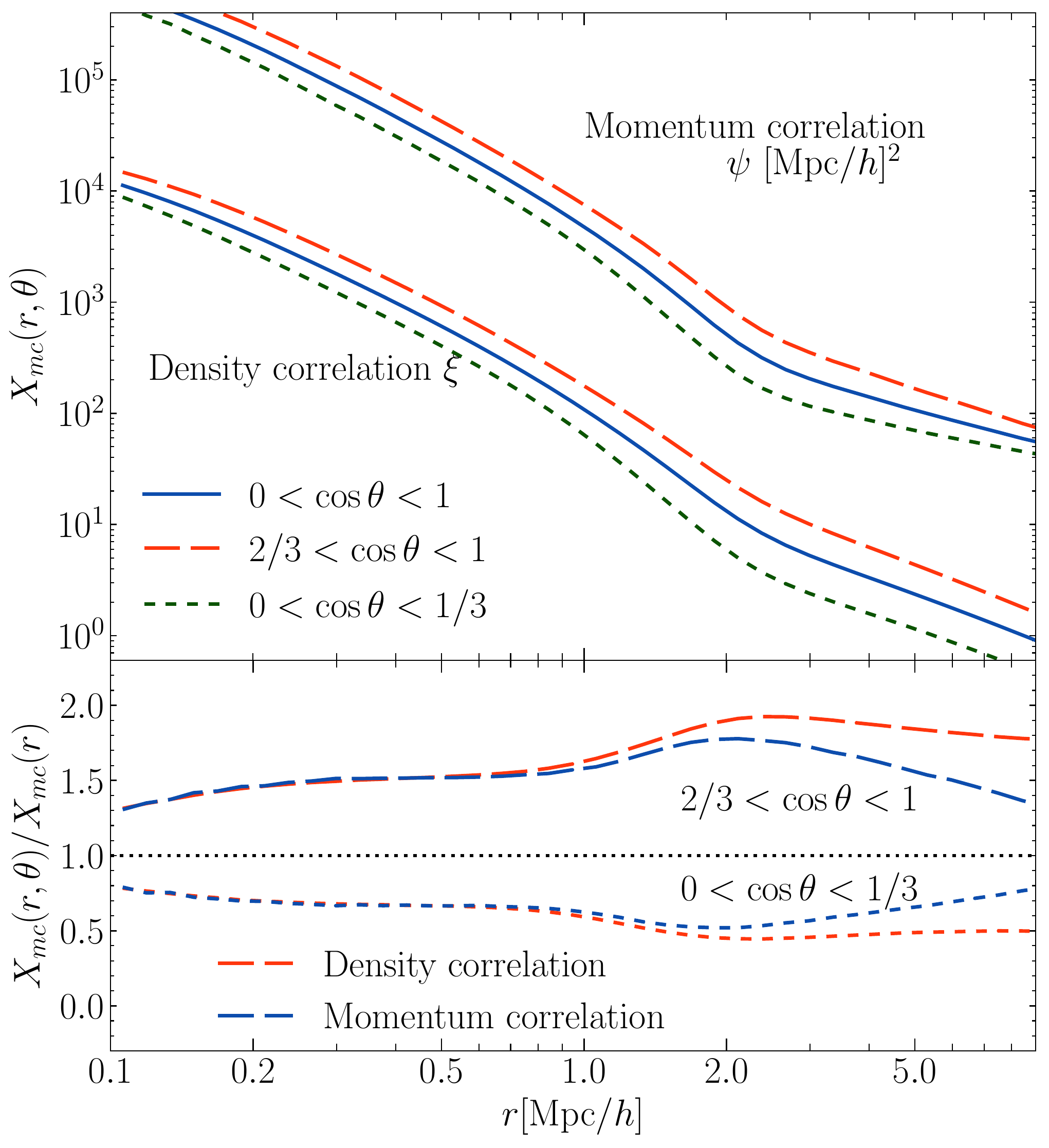}
\caption{Top left: Alignment density correlation (lower lines) and momentum
  correlation (upper lines) of galaxies and clusters. The dashed and
  dotted lines respectively show the correlations parallel and
  perpendicular to the major axes of the clusters, $X_{gc}(r,\theta)$,
  while the solid lines are the conventional, angle-averaged
  statistics, $X_{gc}(r)$, where $X=\{\xi,\psi\}$.  Bottom left: Ratios of
  angle-binned correlations to the conventional ones between galaxies and clusters,
  $X_{gc}(r,\theta)/X_{gc}(r)$. 
  Top right: Same as the top-left panel but the alignment correlation functions of dark matter and clusters, $X_{mc}(r,\theta)$. Bottom right: Ratios of angle-binned correlations to the conventional ones between dark matter and clusters,
  $X_{mc}(r,\theta)/X_{mc}(r)$.  }
\label{fig:xiab}
\end{figure*}

\section{Numerical analysis}\label{sec:analysis}
\subsection{Measurements}
Since one usually considers a correlation
function projected along the line of sight in observational studies,
the splashback features are smeared out to some extent by projection
effects. On the other hand, since we are interested in physical
properties of the splashback radius, we present the alignment density
and velocity correlation statistics in three-dimensional (3D) space.  The alignment
correlation function of the galaxy density and cluster shape can be
measured by
\be 
\xi_{gc}(r,\theta)=\frac{\left\langle D_gD_c\right\rangle}{\left\langle R_gR_c\right\rangle}-1, 
\ee
where 
$\left\langle D_gD_c\right\rangle(r,\theta)$ and $\left\langle R_gR_c\right\rangle(r,\theta)$
are, respectively, the normalized pair counts of the data and their
randoms as functions of separation, $r$, and position angle of
clusters, $\theta$. Note that $\left\langle R_gR_c\right\rangle$ can
be analytically computed since we place the periodic boundary
condition on the simulation box.  The momentum correlation function
can be measured by
\be
\psi_{gc}(r,\theta)=\frac{\left\langle {\bf V}_{g}\cdot{\bf V}_{c}\right\rangle}{\left\langle R_gR_c\right\rangle},
\ee
where 
$\left\langle {\bf V}_{g}\cdot{\bf V}_{c}\right\rangle(r,\theta)$
is the normalized pair count of galaxies and clusters weighted by the
scalar product of their velocities as functions of $r$ and $\theta$.
We also compute the same
statistics, but galaxies are replaced by dark matter as a
density/velocity tracer, namely, $\xi_{mc}(r,\theta)$ and $\psi_{mc}(r,\theta)$.
In the following analysis, we measure these statistics from each of
the 24 realizations and present their means.
We also compute the standard errors from the scatters, 
but we do not show them since the errors are negligibly small. 

The top-left panel of Fig.~\ref{fig:xiab} shows the cluster-galaxy
cross-correlation function binned in the cluster angle $\theta$ for
the density field and that for the velocity field. 
The top-right panel is the same as the top-left panel but shows the cluster-dark matter 
cross-correlation function. 
The bottom panels
present the ratios of the alignment statistic to the conventional
one, $X_{gc}(r,\theta)/X_{gc}(r)$ and $X_{mc}(r,\theta)/X_{mc}(r)$, where $X=\{\xi,\psi\}$.  The
deviation from unity is the evidence of halo-shape alignments.  While
this effect in the density correlation has been extensively studied
both theoretically and observationally
\cite{Faltenbacher:2009,Schneider:2012,Li:2013,Xia:2017}, that in the
momentum correlation is measured by us for the first time.

\begin{figure}[b]
\includegraphics[width=0.45\textwidth,angle=0,clip]{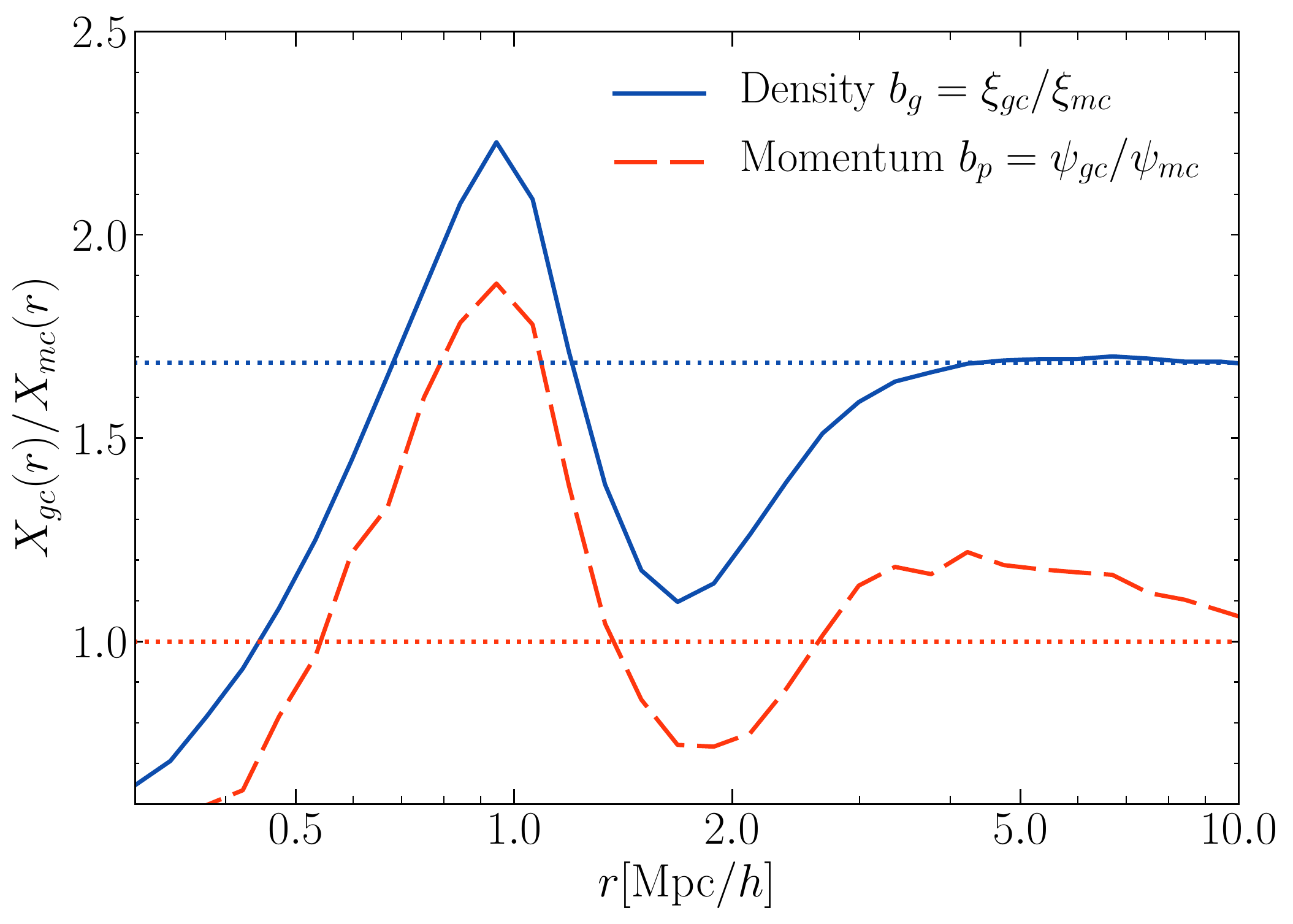}
\caption{Galaxy bias (blue) and momentum bias (red) obtained from the cross-correlation functions, $b_g(r)=\xi_{gc}(r)/\xi_{mc}(r)$ and $b_p(r)=\psi_{gc}(r)/\psi_{mc}(r)$, respectively. 
The blue and red horizontal lines represent the large-scale limit of the galaxy and momentum bias, respectively $b_g=1.69$ and $b_p=1$. 
}
\label{fig:bias}
\end{figure}

Using the cluster-galaxy and cluster-dark matter cross-correlation functions, 
one can determine the galaxy density and momentum biases, respectively, as 
\be
b_g(r)=\frac{\xi_{gc}(r)}{\xi_{mc}(r)}, \ \ \ 
b_p(r)=\frac{\psi_{gc}(r)}{\psi_{mc}(r)}. 
\ee
On sufficiently large scales in which linear perturbation theory is believed to be applicable, 
they approach constants, and particularly we have $b_p=1$ in the absence of the velocity bias. 
See Ref. \cite{Okumura:2012b} for redshift and halo mass dependences of these bias parameters.  
We present the density bias and momentum bias 
in the left and right panels of Fig.~\ref{fig:bias}, respectively. 
The horizontal blue lines are the bias values at the large-scale limit. 
As studied in detail by Ref. \cite{Okumura:2012b}, the density bias approaches the 
linear bias on smaller scales than the momentum bias. 

\begin{figure*}[bt]
\includegraphics[width=0.45\textwidth,angle=0,clip]{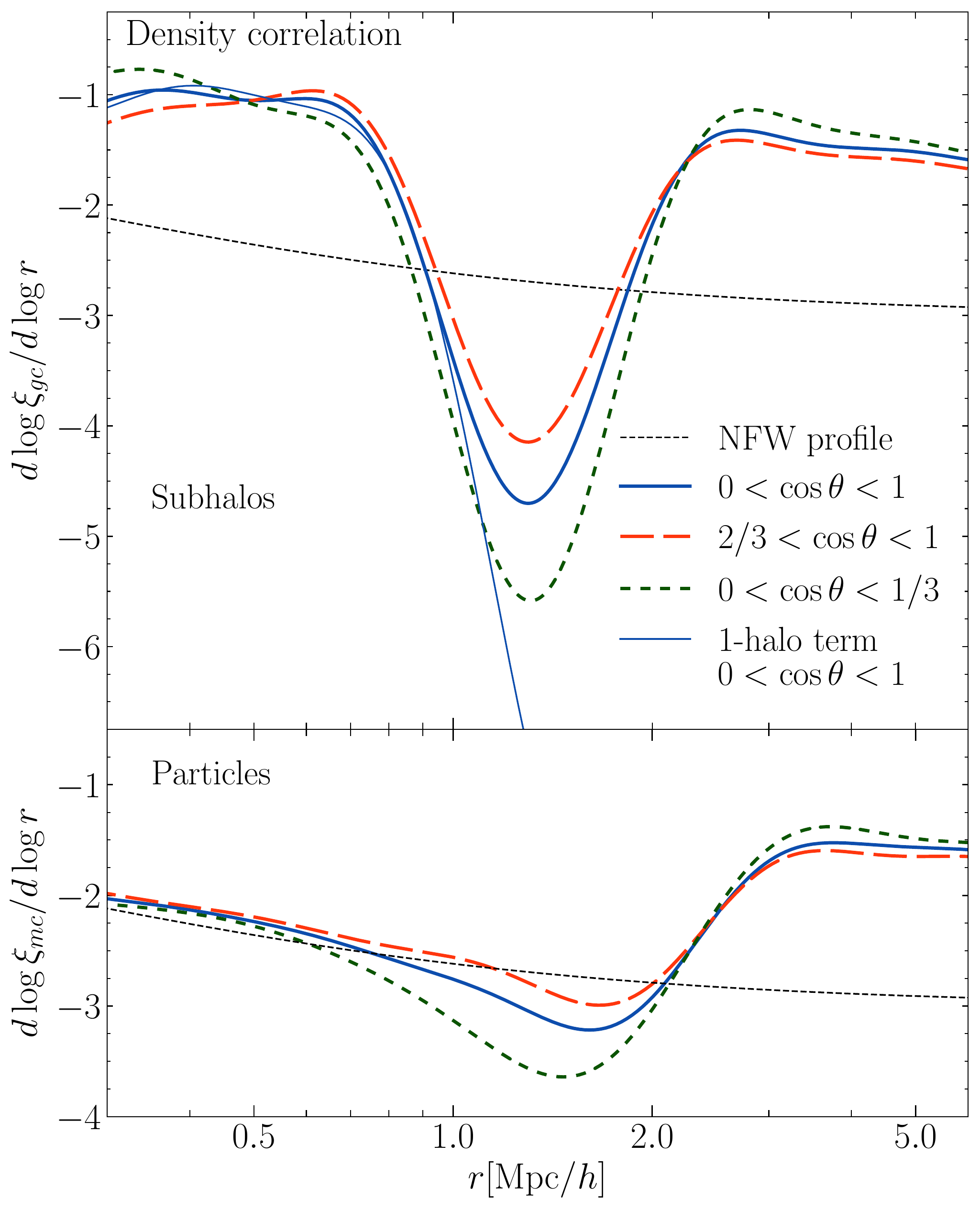}
\includegraphics[width=0.45\textwidth,angle=0,clip]{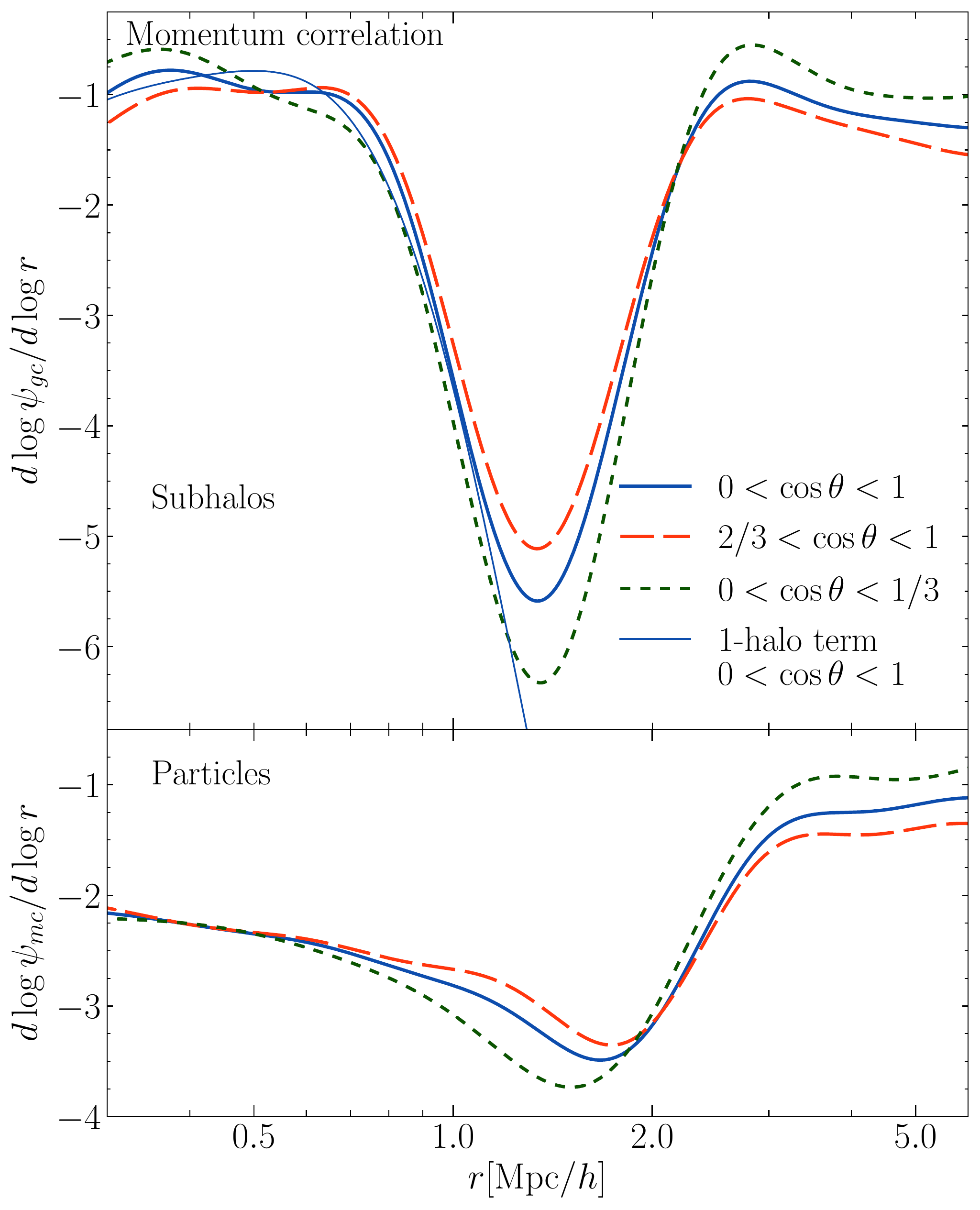}
\caption{Top: Logarithmic slope of the angle-binned galaxy-cluster
  cross-correlation of density (left) and momentum (right). The
  meaning of the line types and colors is the same as the top panel of
  Fig. \ref{fig:xiab}.  The thin blue solid curve is the result of the
  one-halo term, $d\log{X_{gc}^{1h}}/d\log{r}$.  Bottom: Same as the top
  panels, but the density and velocity fields of galaxies are replaced
  by those of dark matter. The black dotted curves in the left panels
  are the NFW profile.  }
\label{fig:xi_splash}
\end{figure*}

\subsection{Splashback features}
On the scale of $x\sim 1\hmpc$, an
abrupt change in the slope of both the density and momentum
correlations can be seen, coinciding with the transition scale between
one-halo and two-halo regimes \cite{Cooray:2002,Schneider:2010}.  As
described in Sec. \ref{sec:intro}, this steepest-slope location is
regarded as a signature of the splashback radius $R_\mathrm{sp}$
\cite{Diemer:2014,Adhikari:2014,More:2015a,Diemer:2017} if the
logarithmic slope there, $\gamma = d\log{\xi_{gc}}/d\log{r}$, is
steeper than the Navarro-Frenk-White (NFW) profile \cite{Navarro:1996}
(See Ref. \cite{Okumura:2015} for an alternative approach to defining the
boundary using the satellite distribution).  The top-left panel of
Fig. \ref{fig:xi_splash} presents $\gamma$, where the derivatives are
computed by interpolating the measured $\xi_{gc}$ with the Gaussian
processes \cite{Rasmussen:2006}.  We used the ``squared
exponential'' kernel and optimized its amplitude and length parameters
to fit $\log{X_{gc}}$ as a function of $\log{r}$ in the range $0.1
\hmpc < x < 10 \hmpc$ \cite{Ambikasaran:2015}.  The steepest slope of
our conventional correlation function reaches $\gamma \simeq -4.5$.
This is significantly steeper than the NFW profile depicted as the black
dotted curve ($\gamma \to -3$) and consistent with the characteristic
splashback properties in $\Lambda$CDM found by
Ref. \cite{Diemer:2014}. Intriguingly, the slopes are shallower and
steeper, respectively, for the alignment correlation parallel and
perpendicular to the major axes of the clusters. This can be
interpreted by the fact that the halo size varies more significantly
along the direction of the major axis (i.e., smaller $\theta$),
leading to a less prominent boundary.

Likewise, we compute the logarithmic derivative for the momentum
correlation, $\gamma_p=d\log{\psi_{gc}}/d\log{r}$, in the top-right
panel of Fig.~\ref{fig:xi_splash}.  It is interesting to note that the
slope approaches $\gamma_p \to -5.5$, steeper than the boundary slope
determined by the density correlation. As discussed by Ref. 
\cite{Lapi:2009} in the context of $\Lambda$CDM structure formation,
the orbital velocity anisotropy is tightly coupled with the
logarithmic density slope around halos and thus expected to be
sensitive to the location of the halo edge \cite{Diemer:2014}, which
physically and sharply separates the multistream intrahalo region
from the outer infall region (see also Ref. \cite{Faltenbacher:2010}).

Another reason for the sharper splashback feature in the momentum
correlation is that it is much less biased than the density
correlation.  Because of the nonlinear, scale-dependent bias, the steepening 
splashback feature probed by the density profile is further smeared
out by the shallow two-halo term (See Ref. \cite{Okumura:2014} for the full
bias dependence on the velocity statistics). To confirm this, we split
the correlation of galaxies and clusters into the correlation with
those inside and outside the halo, $X_{gc}=X_{gc}^{1h}+X_{gc}^{2h}$,
where $X=\{\xi,\psi\}$.  We measure the one-halo terms, $X_{gc}^{1h}$, by cross-correlating 
clusters and their member galaxies identified by the phase-space friends-of-friends technique in the 
R{\footnotesize OCKSTAR} algorithm.
We then take their derivatives, $d\log{X_{gc}^{1h}}/d\log{r}$.  As shown as the blue thin
curves in the top panels of Fig. \ref{fig:xi_splash},
the logarithmic slope profiles in the one-halo regime
probed by the density and momentum correlations are essentially the same, and the difference is less than a few percent around the splashback radius.
This confirms that the differences in
the sharpness of the steepening and
the location of the steepest slope
come from the two-halo terms.

To further see the effect of the bias, we also compute the same
derivatives, but galaxies are replaced by dark matter as a
density/velocity tracer, namely, $d\log{X_{mc}}/d\log{r}$ ($X=\{\xi,\psi\}$).
They are shown in
the lower panels of Fig. \ref{fig:xi_splash}.  As explained in Ref. 
\cite{Mansfield:2017}, the splashback feature in the mass density is
smeared out compared to that traced by subhalos, because of extended
substructures that are abundant in the halo outskirts. However, the
slope is still steeper than that of the NFW profile as depicted by the
black dotted line.

\begin{figure}[b]
\includegraphics[width=0.47\textwidth,angle=0,clip]{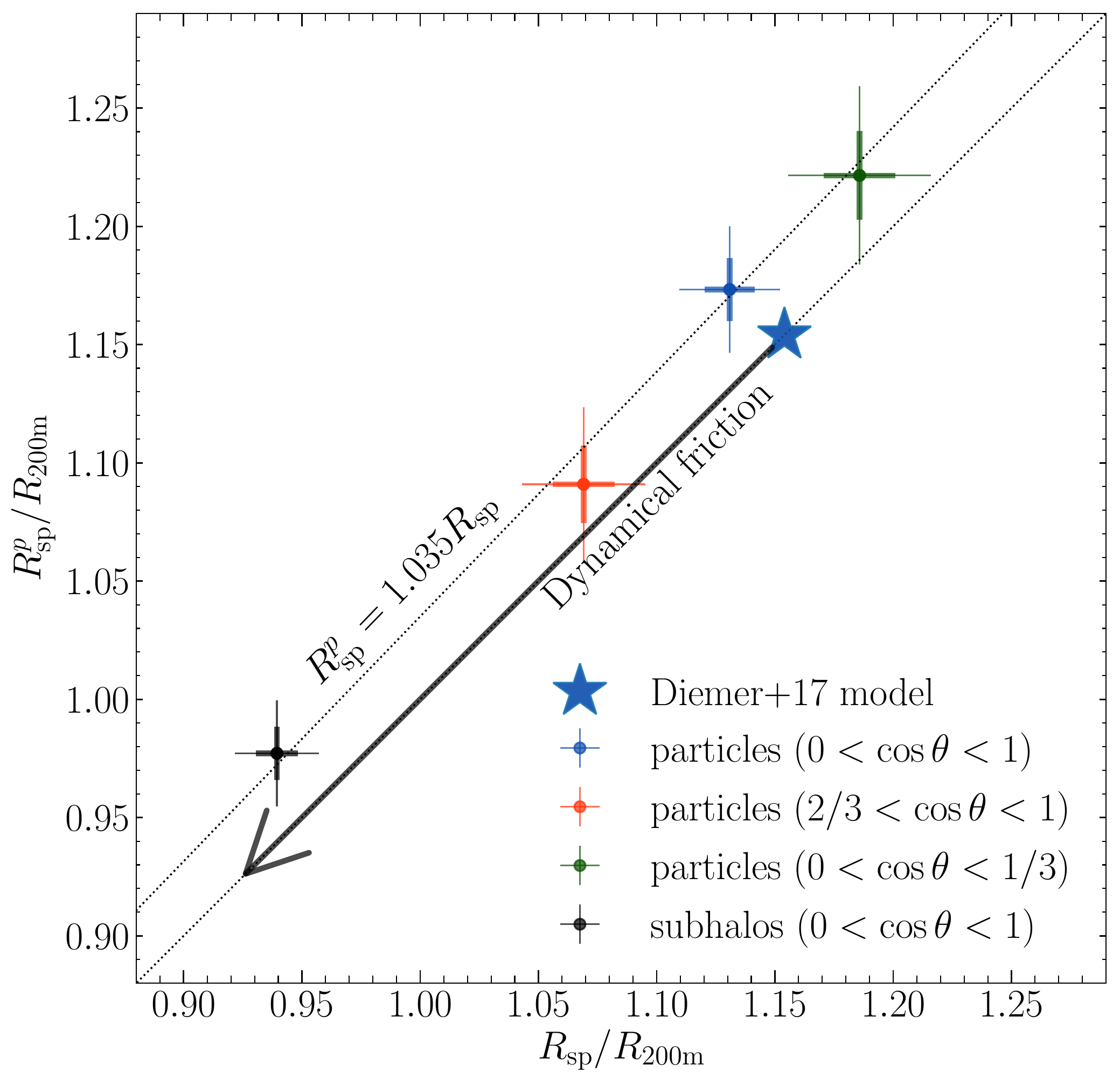}
\caption{Constraints on the splashback radius from the density and
  momentum fields, respectively shown in the horizontal and vertical
  axes. The blue cross is the constraints when the profile is traced
  by all the dark matter particles, while the green and red ones are
  when only the particles parallel and perpendicular to the major axis
  of a halo are used. The thick and thin error bars correspond to
  $1\sigma$ and $2\sigma$ confidence levels, respectively. The black
  cross is the result when the profile is traced by all the
  subhalos. The blue star is a prediction of Ref. \cite{Diemer:2017a}. The
  arrow indicates a range of shifts for the splashback radius due to
  the dynamical friction of subhalos. The upper dotted line represents
  $R_{\rm sp}^p=1.035R_{\rm sp}$.}
\label{fig:r_sp_contour}
\end{figure}

\subsection{Constraints on the splashback radius}
We constrain the splashback radius, $R_\mathrm{sp}$ and $R_\mathrm{sp}^p$, 
the locations of the steepest slopes in the density correlation ($d\gamma/dr |_{r=R_{\rm sp}}=0$) and momentum correlation ($d\gamma_p/dr |_{r=R_{\rm sp}^p}=0$), respectively.  The covariance
error matrix is estimated by the scatter among our 24 realizations.
In Fig. \ref{fig:r_sp_contour}, the blue point with the error bars is
the constraint from the angle-averaged halo-matter cross-correlations.
The blue star represents a model prediction of Ref. \cite{Diemer:2017a},
defined as the 75th percentile of the distribution of particle
apocenters, corresponding approximately to the the radius of steepest
slope in simulated halos \cite{More:2015a}. For our cluster sample
with $\overline{M}_\mathrm{200m}=1.88\times 10^{14} h^{-1}M_\odot$, it
gives $R_\mathrm{sp}=1.15R_{\rm 200m}=1.58\hmpc$, consistent with our
constraints within $1\sigma$.  The green and red points are the
constraints from the correlations parallel and perpendicular to the
halo major axis, respectively.  A clear axis dependence is detected:
$R_\mathrm{sp}$ along and perpendicular to the major axis is
constrained to be higher and lower than the spherically averaged
value.

The resulting locations of $R_\mathrm{sp}$ and $R_\mathrm{sp}^p$
from the subhalo number density (black cross) are significantly
smaller than those from matter density.  This can be interpreted as
the effect of dynamical friction \cite{Adhikari:2016}.  According to
Ref. \cite{Diemer:2017}, $R_\mathrm{sp}$ constrained from the subhalo field
can be smaller than the true value by up to $\sim 20\%$ as indicated
by the arrow in Fig. \ref{fig:r_sp_contour}.  Since we find the
correlation matrix of $R_\mathrm{sp}$ and $R_\mathrm{sp}^p$ is
$0.605$, the momentum correlation indeed provides extra information on
the splashback radius.  Interestingly, the determined $R_\mathrm{sp}$
is $\sim 3.5\%$ smaller than $R_\mathrm{sp}^p$ for all the cases
studied here.
Hence, one can use this tight correlation to infer the value of
$R_\mathrm{sp}^p$ from $R_\mathrm{sp}$ or vice versa.
This slight shift is qualitatively interpreted as follows. 
The momentum correlation is equivalent to the density correlation  
weighted by velocities of tracers. As demonstrated by Ref. \cite{More:2015a},
the infall velocity reaches its (most negative) minimum
at the radius larger than the splashback radius determined by the density profile. 
Thus, the splashback radius determined in 3D space, $R_{\rm sp}$, is 
systematically smaller than that in phase space (six dimensions). 
However, this relation should be tested in more detail 
for different redshifts and halo masses in
future work. 
On the other hand, the fact that the relation $R_{\rm sp}^p=1.035R_{\rm sp}$ holds for both dark matter and galaxies indicates that the galaxy bias does not affect the location of the splashback feature at the level studied in this paper. 

\section{Conclusions}\label{sec:conclusions}
We have proposed a velocity statistic to
investigate detailed properties of the splashback radius of
nonspherical dark-matter halos.  The splashback radius in the
momentum correlation, $R_\mathrm{sp}^p$, has been detected for the
first time from simulations. The feature is even sharper than that in
the density correlation, because it separates distinct infall and
multistream regions of collisionless cold dark matter. 
By measuring the
splashback radius from both density and momentum correlations, we also
demonstrated that the commonly used density statistic yields
$R_\mathrm{sp}$ that is about $3.5\%$ smaller than that expected 
in phase space, $R_\mathrm{sp}^p$.
In other words, the velocity field provides a less biased estimator
to probe the halo boundaries.  Under certain conditions, a
self-interacting dark matter scenario \cite{Spergel:2000} can predict
a $\sim 20\%$ smaller value of the splashback radius in the satellite
distribution, through the drag force acting between dark-matter
particles of subhalos and cluster halos \cite{More:2016}.  The small
shift due to radial infall velocity thus needs to be taken into account in
precise theoretical modeling.  

We found clear dependences of halo
asphericity on splashback features: the density/velocity slopes are
shallower along the halo major axis. We have also determined the
orientation-dependent splashback radius using the cluster-matter
density cross-correlation as well as the momentum cross-correlation. This demonstrates that the collisional
feature of the cold dark matter model can be constrained by precision measurements
of anisotropic splashback features because the halo shape is predicted to be rounder in such a model (see the discussion in Ref. \cite{More:2016}). 

This paper has focused on the alignment statistics only on small,
cluster scales.  In a companion paper \cite{Okumura:2018}, we perform
a comprehensive analysis by extending to larger scales, $x >
100\hmpc$.

\begin{acknowledgments}
We thank Benedikt Diemer for useful comments. We also thank the anonymous referees for many useful suggestions. This work was in part  supported by MEXT Grant-in-Aid for Scientific Research on Innovative Areas (Grants No.~15H05887, No. 15H05893, No. 15K21733, and No. 15H05892). We  acknowledge support from the Ministry of Science and Technology of Taiwan under Grants No. MOST 106-2119-M-001-031-MY3 (T.~O.) and No. 103-2112-M-001-030-MY3 (K.~U.). T.~N. acknowledges financial support from JSPS KAKENHI Grant No. 17K14273 and JST CREST Grant No. JPMJCR1414. K.~O. is supported by Research Fellowships of JSPS for Young Scientists. K.~O. acknowledges financial support from JSPS Grant-in-Aid for JSPS Research Fellow Grant No. JP16J01512. Numerical simulations were carried out on Cray XC30 at the Center for Computational Astrophysics, National Astronomical Observatory of Japan.
\end{acknowledgments}

\bibliography{../../../../refs}

\begin{thebibliography}{62}
\expandafter\ifx\csname natexlab\endcsname\relax\def\natexlab#1{#1}\fi
\expandafter\ifx\csname bibnamefont\endcsname\relax
  \def\bibnamefont#1{#1}\fi
\expandafter\ifx\csname bibfnamefont\endcsname\relax
  \def\bibfnamefont#1{#1}\fi
\expandafter\ifx\csname citenamefont\endcsname\relax
  \def\citenamefont#1{#1}\fi
\expandafter\ifx\csname url\endcsname\relax
  \def\url#1{\texttt{#1}}\fi
\expandafter\ifx\csname urlprefix\endcsname\relax\def\urlprefix{URL }\fi
\providecommand{\bibinfo}[2]{#2}
\providecommand{\eprint}[2][]{\url{#2}}

\bibitem[{\citenamefont{{Cooray} and {Sheth}}(2002)}]{Cooray:2002}
\bibinfo{author}{\bibfnamefont{A.}~\bibnamefont{{Cooray}}} \bibnamefont{and}
  \bibinfo{author}{\bibfnamefont{R.}~\bibnamefont{{Sheth}}},
  \bibinfo{journal}{\physrep} \textbf{\bibinfo{volume}{372}},
  \bibinfo{pages}{1} (\bibinfo{year}{2002})

\bibitem[{\citenamefont{{Mo} et~al.}(2010)\citenamefont{{Mo}, {van den Bosch},
  and {White}}}]{Mo:2010}
\bibinfo{author}{\bibfnamefont{H.}~\bibnamefont{{Mo}}},
  \bibinfo{author}{\bibfnamefont{F.~C.} \bibnamefont{{van den Bosch}}},
  \bibnamefont{and} \bibinfo{author}{\bibfnamefont{S.}~\bibnamefont{{White}}},
  \emph{\bibinfo{title}{{Galaxy Formation and Evolution}}}
  (\bibinfo{year}{2010}).

\bibitem[{\citenamefont{{Diemer} and {Kravtsov}}(2014)}]{Diemer:2014}
\bibinfo{author}{\bibfnamefont{B.}~\bibnamefont{{Diemer}}} \bibnamefont{and}
  \bibinfo{author}{\bibfnamefont{A.~V.} \bibnamefont{{Kravtsov}}},
  \bibinfo{journal}{\apj} \textbf{\bibinfo{volume}{789}}, \bibinfo{eid}{1}
  (\bibinfo{year}{2014})

\bibitem[{\citenamefont{{Adhikari} et~al.}(2014)\citenamefont{{Adhikari},
  {Dalal}, and {Chamberlain}}}]{Adhikari:2014}
\bibinfo{author}{\bibfnamefont{S.}~\bibnamefont{{Adhikari}}},
  \bibinfo{author}{\bibfnamefont{N.}~\bibnamefont{{Dalal}}}, \bibnamefont{and}
  \bibinfo{author}{\bibfnamefont{R.~T.} \bibnamefont{{Chamberlain}}},
  \bibinfo{journal}{\jcap} \textbf{\bibinfo{volume}{11}}, \bibinfo{eid}{019}
  (\bibinfo{year}{2014})

\bibitem[{\citenamefont{{More} et~al.}(2015)\citenamefont{{More}, {Diemer}, and
  {Kravtsov}}}]{More:2015a}
\bibinfo{author}{\bibfnamefont{S.}~\bibnamefont{{More}}},
  \bibinfo{author}{\bibfnamefont{B.}~\bibnamefont{{Diemer}}}, \bibnamefont{and}
  \bibinfo{author}{\bibfnamefont{A.~V.} \bibnamefont{{Kravtsov}}},
  \bibinfo{journal}{\apj} \textbf{\bibinfo{volume}{810}}, \bibinfo{eid}{36}
  (\bibinfo{year}{2015})

\bibitem[{\citenamefont{{Shi}}(2016)}]{Shi:2016}
\bibinfo{author}{\bibfnamefont{X.}~\bibnamefont{{Shi}}},
  \bibinfo{journal}{\mnras} \textbf{\bibinfo{volume}{459}},
  \bibinfo{pages}{3711} (\bibinfo{year}{2016})

\bibitem[{\citenamefont{{Rykoff} et~al.}(2014)\citenamefont{{Rykoff}, {Rozo},
  {Busha}, {Cunha}, {Finoguenov}, {Evrard}, {Hao}, {Koester}, {Leauthaud},
  {Nord} et~al.}}]{Rykoff:2014}
\bibinfo{author}{\bibfnamefont{E.~S.} \bibnamefont{{Rykoff}}},
  \bibinfo{author}{\bibfnamefont{E.}~\bibnamefont{{Rozo}}},
  \bibinfo{author}{\bibfnamefont{M.~T.} \bibnamefont{{Busha}}},
  \bibinfo{author}{\bibfnamefont{C.~E.} \bibnamefont{{Cunha}}},
  \bibinfo{author}{\bibfnamefont{A.}~\bibnamefont{{Finoguenov}}},
  \bibinfo{author}{\bibfnamefont{A.}~\bibnamefont{{Evrard}}},
  \bibinfo{author}{\bibfnamefont{J.}~\bibnamefont{{Hao}}},
  \bibinfo{author}{\bibfnamefont{B.~P.} \bibnamefont{{Koester}}},
  \bibinfo{author}{\bibfnamefont{A.}~\bibnamefont{{Leauthaud}}},
  \bibinfo{author}{\bibfnamefont{B.}~\bibnamefont{{Nord}}},
  \bibnamefont{et~al.}, \bibinfo{journal}{\apj} \textbf{\bibinfo{volume}{785}},
  \bibinfo{eid}{104} (\bibinfo{year}{2014})

\bibitem[{\citenamefont{{Oguri}}(2014)}]{Oguri:2014}
\bibinfo{author}{\bibfnamefont{M.}~\bibnamefont{{Oguri}}},
  \bibinfo{journal}{\mnras} \textbf{\bibinfo{volume}{444}},
  \bibinfo{pages}{147} (\bibinfo{year}{2014})

\bibitem[{\citenamefont{{More} et~al.}(2016)\citenamefont{{More}, {Miyatake},
  {Takada}, {Diemer}, {Kravtsov}, {Dalal}, {More}, {Murata}, {Mandelbaum},
  {Rozo} et~al.}}]{More:2016}
\bibinfo{author}{\bibfnamefont{S.}~\bibnamefont{{More}}},
  \bibinfo{author}{\bibfnamefont{H.}~\bibnamefont{{Miyatake}}},
  \bibinfo{author}{\bibfnamefont{M.}~\bibnamefont{{Takada}}},
  \bibinfo{author}{\bibfnamefont{B.}~\bibnamefont{{Diemer}}},
  \bibinfo{author}{\bibfnamefont{A.~V.} \bibnamefont{{Kravtsov}}},
  \bibinfo{author}{\bibfnamefont{N.~K.} \bibnamefont{{Dalal}}},
  \bibinfo{author}{\bibfnamefont{A.}~\bibnamefont{{More}}},
  \bibinfo{author}{\bibfnamefont{R.}~\bibnamefont{{Murata}}},
  \bibinfo{author}{\bibfnamefont{R.}~\bibnamefont{{Mandelbaum}}},
  \bibinfo{author}{\bibfnamefont{E.}~\bibnamefont{{Rozo}}},
  \bibnamefont{et~al.}, \bibinfo{journal}{\apj} \textbf{\bibinfo{volume}{825}},
  \bibinfo{eid}{39} (\bibinfo{year}{2016})

\bibitem[{\citenamefont{{Umetsu} and {Diemer}}(2017)}]{Umetsu:2017}
\bibinfo{author}{\bibfnamefont{K.}~\bibnamefont{{Umetsu}}} \bibnamefont{and}
  \bibinfo{author}{\bibfnamefont{B.}~\bibnamefont{{Diemer}}},
  \bibinfo{journal}{\apj} \textbf{\bibinfo{volume}{836}}, \bibinfo{eid}{231}
  (\bibinfo{year}{2017})

\bibitem[{\citenamefont{{Baxter} et~al.}(2017)\citenamefont{{Baxter}, {Chang},
  {Jain}, {Adhikari}, {Dalal}, {Kravtsov}, {More}, {Rozo}, {Rykoff}, and
  {Sheth}}}]{Baxter:2017}
\bibinfo{author}{\bibfnamefont{E.}~\bibnamefont{{Baxter}}},
  \bibinfo{author}{\bibfnamefont{C.}~\bibnamefont{{Chang}}},
  \bibinfo{author}{\bibfnamefont{B.}~\bibnamefont{{Jain}}},
  \bibinfo{author}{\bibfnamefont{S.}~\bibnamefont{{Adhikari}}},
  \bibinfo{author}{\bibfnamefont{N.}~\bibnamefont{{Dalal}}},
  \bibinfo{author}{\bibfnamefont{A.}~\bibnamefont{{Kravtsov}}},
  \bibinfo{author}{\bibfnamefont{S.}~\bibnamefont{{More}}},
  \bibinfo{author}{\bibfnamefont{E.}~\bibnamefont{{Rozo}}},
  \bibinfo{author}{\bibfnamefont{E.}~\bibnamefont{{Rykoff}}}, \bibnamefont{and}
  \bibinfo{author}{\bibfnamefont{R.~K.} \bibnamefont{{Sheth}}},
  \bibinfo{journal}{\apj} \textbf{\bibinfo{volume}{841}}, \bibinfo{eid}{18}
  (\bibinfo{year}{2017})

\bibitem[{\citenamefont{{Chang} et~al.}(2017)\citenamefont{{Chang}, {Baxter},
  {Jain}, {S{\'a}nchez}, {Adhikari}, {Varga}, {Fang}, {Rozo}, {Rykoff},
  {Kravtsov} et~al.}}]{Chang:2017}
\bibinfo{author}{\bibfnamefont{C.}~\bibnamefont{{Chang}}},
  \bibinfo{author}{\bibfnamefont{E.}~\bibnamefont{{Baxter}}},
  \bibinfo{author}{\bibfnamefont{B.}~\bibnamefont{{Jain}}},
  \bibinfo{author}{\bibfnamefont{C.}~\bibnamefont{{S{\'a}nchez}}},
  \bibinfo{author}{\bibfnamefont{S.}~\bibnamefont{{Adhikari}}},
  \bibinfo{author}{\bibfnamefont{T.~N.} \bibnamefont{{Varga}}},
  \bibinfo{author}{\bibfnamefont{Y.}~\bibnamefont{{Fang}}},
  \bibinfo{author}{\bibfnamefont{E.}~\bibnamefont{{Rozo}}},
  \bibinfo{author}{\bibfnamefont{E.~S.} \bibnamefont{{Rykoff}}},
  \bibinfo{author}{\bibfnamefont{A.}~\bibnamefont{{Kravtsov}}},
  \bibnamefont{et~al.}, \bibinfo{journal}{ArXiv e-prints}
  (\bibinfo{year}{2017}), \eprint{1710.06808}.

\bibitem[{\citenamefont{{Busch} and {White}}(2017)}]{Busch:2017}
\bibinfo{author}{\bibfnamefont{P.}~\bibnamefont{{Busch}}} \bibnamefont{and}
  \bibinfo{author}{\bibfnamefont{S.~D.~M.} \bibnamefont{{White}}},
  \bibinfo{journal}{\mnras} \textbf{\bibinfo{volume}{470}},
  \bibinfo{pages}{4767} (\bibinfo{year}{2017})

\bibitem[{\citenamefont{{Adhikari} et~al.}(2016)\citenamefont{{Adhikari},
  {Dalal}, and {Clampitt}}}]{Adhikari:2016}
\bibinfo{author}{\bibfnamefont{S.}~\bibnamefont{{Adhikari}}},
  \bibinfo{author}{\bibfnamefont{N.}~\bibnamefont{{Dalal}}}, \bibnamefont{and}
  \bibinfo{author}{\bibfnamefont{J.}~\bibnamefont{{Clampitt}}},
  \bibinfo{journal}{\jcap} \textbf{\bibinfo{volume}{7}}, \bibinfo{eid}{022}
  (\bibinfo{year}{2016})

\bibitem[{\citenamefont{{Diemer}}(2017)}]{Diemer:2017}
\bibinfo{author}{\bibfnamefont{B.}~\bibnamefont{{Diemer}}},
  \bibinfo{journal}{\apjs} \textbf{\bibinfo{volume}{231}}, \bibinfo{eid}{5}
  (\bibinfo{year}{2017})

\bibitem[{\citenamefont{{Diemer} et~al.}(2017)\citenamefont{{Diemer},
  {Mansfield}, {Kravtsov}, and {More}}}]{Diemer:2017a}
\bibinfo{author}{\bibfnamefont{B.}~\bibnamefont{{Diemer}}},
  \bibinfo{author}{\bibfnamefont{P.}~\bibnamefont{{Mansfield}}},
  \bibinfo{author}{\bibfnamefont{A.~V.} \bibnamefont{{Kravtsov}}},
  \bibnamefont{and} \bibinfo{author}{\bibfnamefont{S.}~\bibnamefont{{More}}},
  \bibinfo{journal}{\apj} \textbf{\bibinfo{volume}{843}}, \bibinfo{eid}{140}
  (\bibinfo{year}{2017})

\bibitem[{\citenamefont{{Mansfield} et~al.}(2017)\citenamefont{{Mansfield},
  {Kravtsov}, and {Diemer}}}]{Mansfield:2017}
\bibinfo{author}{\bibfnamefont{P.}~\bibnamefont{{Mansfield}}},
  \bibinfo{author}{\bibfnamefont{A.~V.} \bibnamefont{{Kravtsov}}},
  \bibnamefont{and} \bibinfo{author}{\bibfnamefont{B.}~\bibnamefont{{Diemer}}},
  \bibinfo{journal}{\apj} \textbf{\bibinfo{volume}{841}}, \bibinfo{eid}{34}
  (\bibinfo{year}{2017})

\bibitem[{\citenamefont{{Diemand} and {Kuhlen}}(2008)}]{Diemand:2008}
\bibinfo{author}{\bibfnamefont{J.}~\bibnamefont{{Diemand}}} \bibnamefont{and}
  \bibinfo{author}{\bibfnamefont{M.}~\bibnamefont{{Kuhlen}}},
  \bibinfo{journal}{\apjl} \textbf{\bibinfo{volume}{680}}, \bibinfo{eid}{L25}
  (\bibinfo{year}{2008})

\bibitem[{\citenamefont{{Lemze} et~al.}(2009)\citenamefont{{Lemze},
  {Broadhurst}, {Rephaeli}, {Barkana}, and {Umetsu}}}]{Lemze:2009}
\bibinfo{author}{\bibfnamefont{D.}~\bibnamefont{{Lemze}}},
  \bibinfo{author}{\bibfnamefont{T.}~\bibnamefont{{Broadhurst}}},
  \bibinfo{author}{\bibfnamefont{Y.}~\bibnamefont{{Rephaeli}}},
  \bibinfo{author}{\bibfnamefont{R.}~\bibnamefont{{Barkana}}},
  \bibnamefont{and} \bibinfo{author}{\bibfnamefont{K.}~\bibnamefont{{Umetsu}}},
  \bibinfo{journal}{\apj} \textbf{\bibinfo{volume}{701}}, \bibinfo{pages}{1336}
  (\bibinfo{year}{2009})

\bibitem[{\citenamefont{{Vogelsberger}
  et~al.}(2011)\citenamefont{{Vogelsberger}, {Mohayaee}, and
  {White}}}]{Vogelsberger:2011}
\bibinfo{author}{\bibfnamefont{M.}~\bibnamefont{{Vogelsberger}}},
  \bibinfo{author}{\bibfnamefont{R.}~\bibnamefont{{Mohayaee}}},
  \bibnamefont{and} \bibinfo{author}{\bibfnamefont{S.~D.~M.}
  \bibnamefont{{White}}}, \bibinfo{journal}{\mnras}
  \textbf{\bibinfo{volume}{414}}, \bibinfo{pages}{3044} (\bibinfo{year}{2011})

\bibitem[{\citenamefont{{Rines} et~al.}(2013)\citenamefont{{Rines}, {Geller},
  {Diaferio}, and {Kurtz}}}]{Rines:2013}
\bibinfo{author}{\bibfnamefont{K.}~\bibnamefont{{Rines}}},
  \bibinfo{author}{\bibfnamefont{M.~J.} \bibnamefont{{Geller}}},
  \bibinfo{author}{\bibfnamefont{A.}~\bibnamefont{{Diaferio}}},
  \bibnamefont{and} \bibinfo{author}{\bibfnamefont{M.~J.}
  \bibnamefont{{Kurtz}}}, \bibinfo{journal}{\apj}
  \textbf{\bibinfo{volume}{767}}, \bibinfo{eid}{15} (\bibinfo{year}{2013})

\bibitem[{\citenamefont{{Svensmark} et~al.}(2015)\citenamefont{{Svensmark},
  {Wojtak}, and {Hansen}}}]{Svensmark:2015}
\bibinfo{author}{\bibfnamefont{J.}~\bibnamefont{{Svensmark}}},
  \bibinfo{author}{\bibfnamefont{R.}~\bibnamefont{{Wojtak}}}, \bibnamefont{and}
  \bibinfo{author}{\bibfnamefont{S.~H.} \bibnamefont{{Hansen}}},
  \bibinfo{journal}{\mnras} \textbf{\bibinfo{volume}{448}},
  \bibinfo{pages}{1644} (\bibinfo{year}{2015})

\bibitem[{\citenamefont{{Hayashi} and {White}}(2008)}]{Hayashi:2008}
\bibinfo{author}{\bibfnamefont{E.}~\bibnamefont{{Hayashi}}} \bibnamefont{and}
  \bibinfo{author}{\bibfnamefont{S.~D.~M.} \bibnamefont{{White}}},
  \bibinfo{journal}{\mnras} \textbf{\bibinfo{volume}{388}}, \bibinfo{pages}{2}
  (\bibinfo{year}{2008})

\bibitem[{\citenamefont{{Paz} et~al.}(2008)\citenamefont{{Paz}, {Stasyszyn},
  and {Padilla}}}]{Paz:2008}
\bibinfo{author}{\bibfnamefont{D.~J.} \bibnamefont{{Paz}}},
  \bibinfo{author}{\bibfnamefont{F.}~\bibnamefont{{Stasyszyn}}},
  \bibnamefont{and} \bibinfo{author}{\bibfnamefont{N.~D.}
  \bibnamefont{{Padilla}}}, \bibinfo{journal}{\mnras}
  \textbf{\bibinfo{volume}{389}}, \bibinfo{pages}{1127} (\bibinfo{year}{2008})

\bibitem[{\citenamefont{{Faltenbacher}
  et~al.}(2009)\citenamefont{{Faltenbacher}, {Li}, {White}, {Jing}, {Mao}, and
  {Wang}}}]{Faltenbacher:2009}
\bibinfo{author}{\bibfnamefont{A.}~\bibnamefont{{Faltenbacher}}},
  \bibinfo{author}{\bibfnamefont{C.}~\bibnamefont{{Li}}},
  \bibinfo{author}{\bibfnamefont{S.~D.~M.} \bibnamefont{{White}}},
  \bibinfo{author}{\bibfnamefont{Y.-P.} \bibnamefont{{Jing}}},
  \bibinfo{author}{\bibfnamefont{S.}~\bibnamefont{{Mao}}}, \bibnamefont{and}
  \bibinfo{author}{\bibfnamefont{J.}~\bibnamefont{{Wang}}},
  \bibinfo{journal}{Research in Astronomy and Astrophysics}
  \textbf{\bibinfo{volume}{9}}, \bibinfo{pages}{41} (\bibinfo{year}{2009})

\bibitem[{\citenamefont{{Osato} et~al.}(2018)\citenamefont{{Osato},
  {Nishimichi}, {Oguri}, {Takada}, and {Okumura}}}]{Osato:2018}
\bibinfo{author}{\bibfnamefont{K.}~\bibnamefont{{Osato}}},
  \bibinfo{author}{\bibfnamefont{T.}~\bibnamefont{{Nishimichi}}},
  \bibinfo{author}{\bibfnamefont{M.}~\bibnamefont{{Oguri}}},
  \bibinfo{author}{\bibfnamefont{M.}~\bibnamefont{{Takada}}}, \bibnamefont{and}
  \bibinfo{author}{\bibfnamefont{T.}~\bibnamefont{{Okumura}}},
  \bibinfo{journal}{\mnras} \textbf{\bibinfo{volume}{477}},
  \bibinfo{pages}{2141} (\bibinfo{year}{2018})

\bibitem[{\citenamefont{{Catelan} et~al.}(2001)\citenamefont{{Catelan},
  {Kamionkowski}, and {Blandford}}}]{Catelan:2001}
\bibinfo{author}{\bibfnamefont{P.}~\bibnamefont{{Catelan}}},
  \bibinfo{author}{\bibfnamefont{M.}~\bibnamefont{{Kamionkowski}}},
  \bibnamefont{and} \bibinfo{author}{\bibfnamefont{R.~D.}
  \bibnamefont{{Blandford}}}, \bibinfo{journal}{\mnras}
  \textbf{\bibinfo{volume}{320}}, \bibinfo{pages}{L7} (\bibinfo{year}{2001})

\bibitem[{\citenamefont{{Hirata} and {Seljak}}(2004)}]{Hirata:2004}
\bibinfo{author}{\bibfnamefont{C.~M.} \bibnamefont{{Hirata}}} \bibnamefont{and}
  \bibinfo{author}{\bibfnamefont{U.}~\bibnamefont{{Seljak}}},
  \bibinfo{journal}{\prd} \textbf{\bibinfo{volume}{70}}, \bibinfo{eid}{063526}
  (\bibinfo{year}{2004})

\bibitem[{\citenamefont{{Mandelbaum} et~al.}(2006)\citenamefont{{Mandelbaum},
  {Hirata}, {Ishak}, {Seljak}, and {Brinkmann}}}]{Mandelbaum:2006}
\bibinfo{author}{\bibfnamefont{R.}~\bibnamefont{{Mandelbaum}}},
  \bibinfo{author}{\bibfnamefont{C.~M.} \bibnamefont{{Hirata}}},
  \bibinfo{author}{\bibfnamefont{M.}~\bibnamefont{{Ishak}}},
  \bibinfo{author}{\bibfnamefont{U.}~\bibnamefont{{Seljak}}}, \bibnamefont{and}
  \bibinfo{author}{\bibfnamefont{J.}~\bibnamefont{{Brinkmann}}},
  \bibinfo{journal}{\mnras} \textbf{\bibinfo{volume}{367}},
  \bibinfo{pages}{611} (\bibinfo{year}{2006})

\bibitem[{\citenamefont{{Okumura} et~al.}(2009)\citenamefont{{Okumura}, {Jing},
  and {Li}}}]{Okumura:2009}
\bibinfo{author}{\bibfnamefont{T.}~\bibnamefont{{Okumura}}},
  \bibinfo{author}{\bibfnamefont{Y.~P.} \bibnamefont{{Jing}}},
  \bibnamefont{and} \bibinfo{author}{\bibfnamefont{C.}~\bibnamefont{{Li}}},
  \bibinfo{journal}{\apj} \textbf{\bibinfo{volume}{694}}, \bibinfo{pages}{214}
  (\bibinfo{year}{2009})

\bibitem[{\citenamefont{{Sch{\"a}fer}}(2009)}]{Schafer:2009}
\bibinfo{author}{\bibfnamefont{B.~M.} \bibnamefont{{Sch{\"a}fer}}},
  \bibinfo{journal}{International Journal of Modern Physics D}
  \textbf{\bibinfo{volume}{18}}, \bibinfo{pages}{173} (\bibinfo{year}{2009})

\bibitem[{\citenamefont{{Troxel} and {Ishak}}(2015)}]{Troxel:2015}
\bibinfo{author}{\bibfnamefont{M.~A.} \bibnamefont{{Troxel}}} \bibnamefont{and}
  \bibinfo{author}{\bibfnamefont{M.}~\bibnamefont{{Ishak}}},
  \bibinfo{journal}{\physrep} \textbf{\bibinfo{volume}{558}},
  \bibinfo{pages}{1} (\bibinfo{year}{2015})

\bibitem[{\citenamefont{{Joachimi} et~al.}(2015)\citenamefont{{Joachimi},
  {Cacciato}, {Kitching}, {Leonard}, {Mandelbaum}, {Sch{\"a}fer}, {Sif{\'o}n},
  {Hoekstra}, {Kiessling}, {Kirk} et~al.}}]{Joachimi:2015}
\bibinfo{author}{\bibfnamefont{B.}~\bibnamefont{{Joachimi}}},
  \bibinfo{author}{\bibfnamefont{M.}~\bibnamefont{{Cacciato}}},
  \bibinfo{author}{\bibfnamefont{T.~D.} \bibnamefont{{Kitching}}},
  \bibinfo{author}{\bibfnamefont{A.}~\bibnamefont{{Leonard}}},
  \bibinfo{author}{\bibfnamefont{R.}~\bibnamefont{{Mandelbaum}}},
  \bibinfo{author}{\bibfnamefont{B.~M.} \bibnamefont{{Sch{\"a}fer}}},
  \bibinfo{author}{\bibfnamefont{C.}~\bibnamefont{{Sif{\'o}n}}},
  \bibinfo{author}{\bibfnamefont{H.}~\bibnamefont{{Hoekstra}}},
  \bibinfo{author}{\bibfnamefont{A.}~\bibnamefont{{Kiessling}}},
  \bibinfo{author}{\bibfnamefont{D.}~\bibnamefont{{Kirk}}},
  \bibnamefont{et~al.}, \bibinfo{journal}{\ssr} \textbf{\bibinfo{volume}{193}},
  \bibinfo{pages}{1} (\bibinfo{year}{2015})

\bibitem[{\citenamefont{{van Uitert} and {Joachimi}}(2017)}]{van-Uitert:2017}
\bibinfo{author}{\bibfnamefont{E.}~\bibnamefont{{van Uitert}}}
  \bibnamefont{and}
  \bibinfo{author}{\bibfnamefont{B.}~\bibnamefont{{Joachimi}}},
  \bibinfo{journal}{\mnras} \textbf{\bibinfo{volume}{468}},
  \bibinfo{pages}{4502} (\bibinfo{year}{2017})

\bibitem[{\citenamefont{{Okumura} and {Jing}}(2009)}]{Okumura:2009a}
\bibinfo{author}{\bibfnamefont{T.}~\bibnamefont{{Okumura}}} \bibnamefont{and}
  \bibinfo{author}{\bibfnamefont{Y.~P.} \bibnamefont{{Jing}}},
  \bibinfo{journal}{\apjl} \textbf{\bibinfo{volume}{694}}, \bibinfo{pages}{L83}
  (\bibinfo{year}{2009})

\bibitem[{\citenamefont{{Blazek} et~al.}(2011)\citenamefont{{Blazek},
  {McQuinn}, and {Seljak}}}]{Blazek:2011}
\bibinfo{author}{\bibfnamefont{J.}~\bibnamefont{{Blazek}}},
  \bibinfo{author}{\bibfnamefont{M.}~\bibnamefont{{McQuinn}}},
  \bibnamefont{and} \bibinfo{author}{\bibfnamefont{U.}~\bibnamefont{{Seljak}}},
  \bibinfo{journal}{\jcap} \textbf{\bibinfo{volume}{5}}, \bibinfo{pages}{10}
  (\bibinfo{year}{2011})

\bibitem[{\citenamefont{{Gorski}}(1988)}]{Gorski:1988}
\bibinfo{author}{\bibfnamefont{K.}~\bibnamefont{{Gorski}}},
  \bibinfo{journal}{\apjl} \textbf{\bibinfo{volume}{332}}, \bibinfo{pages}{L7}
  (\bibinfo{year}{1988}).

\bibitem[{\citenamefont{{Fisher}}(1995)}]{Fisher:1995}
\bibinfo{author}{\bibfnamefont{K.~B.} \bibnamefont{{Fisher}}},
  \bibinfo{journal}{\apj} \textbf{\bibinfo{volume}{448}}, \bibinfo{pages}{494}
  (\bibinfo{year}{1995})

\bibitem[{\citenamefont{{Strauss} and {Willick}}(1995)}]{Strauss:1995}
\bibinfo{author}{\bibfnamefont{M.~A.} \bibnamefont{{Strauss}}}
  \bibnamefont{and} \bibinfo{author}{\bibfnamefont{J.~A.}
  \bibnamefont{{Willick}}}, \bibinfo{journal}{\physrep}
  \textbf{\bibinfo{volume}{261}}, \bibinfo{pages}{271} (\bibinfo{year}{1995})

\bibitem[{\citenamefont{{Okumura} et~al.}(2014)\citenamefont{{Okumura},
  {Seljak}, {Vlah}, and {Desjacques}}}]{Okumura:2014}
\bibinfo{author}{\bibfnamefont{T.}~\bibnamefont{{Okumura}}},
  \bibinfo{author}{\bibfnamefont{U.}~\bibnamefont{{Seljak}}},
  \bibinfo{author}{\bibfnamefont{Z.}~\bibnamefont{{Vlah}}}, \bibnamefont{and}
  \bibinfo{author}{\bibfnamefont{V.}~\bibnamefont{{Desjacques}}},
  \bibinfo{journal}{\jcap} \textbf{\bibinfo{volume}{5}}, \bibinfo{eid}{003}
  (\bibinfo{year}{2014})

\bibitem[{\citenamefont{{Sugiyama} et~al.}(2016)\citenamefont{{Sugiyama},
  {Okumura}, and {Spergel}}}]{Sugiyama:2016}
\bibinfo{author}{\bibfnamefont{N.~S.} \bibnamefont{{Sugiyama}}},
  \bibinfo{author}{\bibfnamefont{T.}~\bibnamefont{{Okumura}}},
  \bibnamefont{and} \bibinfo{author}{\bibfnamefont{D.~N.}
  \bibnamefont{{Spergel}}}, \bibinfo{journal}{\jcap}
  \textbf{\bibinfo{volume}{7}}, \bibinfo{eid}{001} (\bibinfo{year}{2016})

\bibitem[{\citenamefont{{Nishimichi, et al.}}(2018)}]{Nishimichi:2018}
\bibinfo{author}{\bibfnamefont{T.}~\bibnamefont{{Nishimichi, et al.}}},
  \bibinfo{journal}{in preparation}  (\bibinfo{year}{2018}).

\bibitem[{\citenamefont{{Behroozi} et~al.}(2013)\citenamefont{{Behroozi},
  {Wechsler}, and {Wu}}}]{Behroozi:2013}
\bibinfo{author}{\bibfnamefont{P.~S.} \bibnamefont{{Behroozi}}},
  \bibinfo{author}{\bibfnamefont{R.~H.} \bibnamefont{{Wechsler}}},
  \bibnamefont{and} \bibinfo{author}{\bibfnamefont{H.-Y.} \bibnamefont{{Wu}}},
  \bibinfo{journal}{\apj} \textbf{\bibinfo{volume}{762}}, \bibinfo{eid}{109}
  (\bibinfo{year}{2013})

\bibitem[{\citenamefont{{Zheng} et~al.}(2005)\citenamefont{{Zheng}, {Berlind},
  {Weinberg}, {Benson}, {Baugh}, {Cole}, {Dave}, {Frenk}, {Katz}, and
  {Lacey}}}]{Zheng:2005}
\bibinfo{author}{\bibfnamefont{Z.}~\bibnamefont{{Zheng}}},
  \bibinfo{author}{\bibfnamefont{A.~A.} \bibnamefont{{Berlind}}},
  \bibinfo{author}{\bibfnamefont{D.~H.} \bibnamefont{{Weinberg}}},
  \bibinfo{author}{\bibfnamefont{A.~J.} \bibnamefont{{Benson}}},
  \bibinfo{author}{\bibfnamefont{C.~M.} \bibnamefont{{Baugh}}},
  \bibinfo{author}{\bibfnamefont{S.}~\bibnamefont{{Cole}}},
  \bibinfo{author}{\bibfnamefont{R.}~\bibnamefont{{Dave}}},
  \bibinfo{author}{\bibfnamefont{C.~S.} \bibnamefont{{Frenk}}},
  \bibinfo{author}{\bibfnamefont{N.}~\bibnamefont{{Katz}}}, \bibnamefont{and}
  \bibinfo{author}{\bibfnamefont{C.~G.} \bibnamefont{{Lacey}}},
  \bibinfo{journal}{\apj} \textbf{\bibinfo{volume}{633}}, \bibinfo{pages}{791}
  (\bibinfo{year}{2005})

\bibitem[{\citenamefont{{Parejko} et~al.}(2013)\citenamefont{{Parejko},
  {Sunayama}, {Padmanabhan}, {Wake}, {Berlind}, {Bizyaev}, {Blanton}, {Bolton},
  {van den Bosch}, {Brinkmann} et~al.}}]{Parejko:2013}
\bibinfo{author}{\bibfnamefont{J.~K.} \bibnamefont{{Parejko}}},
  \bibinfo{author}{\bibfnamefont{T.}~\bibnamefont{{Sunayama}}},
  \bibinfo{author}{\bibfnamefont{N.}~\bibnamefont{{Padmanabhan}}},
  \bibinfo{author}{\bibfnamefont{D.~A.} \bibnamefont{{Wake}}},
  \bibinfo{author}{\bibfnamefont{A.~A.} \bibnamefont{{Berlind}}},
  \bibinfo{author}{\bibfnamefont{D.}~\bibnamefont{{Bizyaev}}},
  \bibinfo{author}{\bibfnamefont{M.}~\bibnamefont{{Blanton}}},
  \bibinfo{author}{\bibfnamefont{A.~S.} \bibnamefont{{Bolton}}},
  \bibinfo{author}{\bibfnamefont{F.}~\bibnamefont{{van den Bosch}}},
  \bibinfo{author}{\bibfnamefont{J.}~\bibnamefont{{Brinkmann}}},
  \bibnamefont{et~al.}, \bibinfo{journal}{\mnras}
  \textbf{\bibinfo{volume}{429}}, \bibinfo{pages}{98} (\bibinfo{year}{2013})

\bibitem[{\citenamefont{{Nishimichi} and {Oka}}(2014)}]{Nishimichi:2014}
\bibinfo{author}{\bibfnamefont{T.}~\bibnamefont{{Nishimichi}}}
  \bibnamefont{and} \bibinfo{author}{\bibfnamefont{A.}~\bibnamefont{{Oka}}},
  \bibinfo{journal}{\mnras} \textbf{\bibinfo{volume}{444}},
  \bibinfo{pages}{1400} (\bibinfo{year}{2014})

\bibitem[{\citenamefont{{Okumura}
  et~al.}(2017{\natexlab{a}})\citenamefont{{Okumura}, {Takada}, {More}, and
  {Masaki}}}]{Okumura:2017}
\bibinfo{author}{\bibfnamefont{T.}~\bibnamefont{{Okumura}}},
  \bibinfo{author}{\bibfnamefont{M.}~\bibnamefont{{Takada}}},
  \bibinfo{author}{\bibfnamefont{S.}~\bibnamefont{{More}}}, \bibnamefont{and}
  \bibinfo{author}{\bibfnamefont{S.}~\bibnamefont{{Masaki}}},
  \bibinfo{journal}{\mnras} \textbf{\bibinfo{volume}{469}},
  \bibinfo{pages}{459} (\bibinfo{year}{2017}{\natexlab{a}})

\bibitem[{\citenamefont{{Jing} and {Suto}}(2002)}]{Jing:2002a}
\bibinfo{author}{\bibfnamefont{Y.~P.} \bibnamefont{{Jing}}} \bibnamefont{and}
  \bibinfo{author}{\bibfnamefont{Y.}~\bibnamefont{{Suto}}},
  \bibinfo{journal}{\apj} \textbf{\bibinfo{volume}{574}}, \bibinfo{pages}{538}
  (\bibinfo{year}{2002})

\bibitem[{\citenamefont{{Schneider} et~al.}(2012)\citenamefont{{Schneider},
  {Frenk}, and {Cole}}}]{Schneider:2012}
\bibinfo{author}{\bibfnamefont{M.~D.} \bibnamefont{{Schneider}}},
  \bibinfo{author}{\bibfnamefont{C.~S.} \bibnamefont{{Frenk}}},
  \bibnamefont{and} \bibinfo{author}{\bibfnamefont{S.}~\bibnamefont{{Cole}}},
  \bibinfo{journal}{\jcap} \textbf{\bibinfo{volume}{5}}, \bibinfo{eid}{030}
  (\bibinfo{year}{2012})

\bibitem[{\citenamefont{{Li} et~al.}(2013)\citenamefont{{Li}, {Jing},
  {Faltenbacher}, and {Wang}}}]{Li:2013}
\bibinfo{author}{\bibfnamefont{C.}~\bibnamefont{{Li}}},
  \bibinfo{author}{\bibfnamefont{Y.~P.} \bibnamefont{{Jing}}},
  \bibinfo{author}{\bibfnamefont{A.}~\bibnamefont{{Faltenbacher}}},
  \bibnamefont{and} \bibinfo{author}{\bibfnamefont{J.}~\bibnamefont{{Wang}}},
  \bibinfo{journal}{\apjl} \textbf{\bibinfo{volume}{770}}, \bibinfo{eid}{L12}
  (\bibinfo{year}{2013})

\bibitem[{\citenamefont{{Xia} et~al.}(2017)\citenamefont{{Xia}, {Kang}, {Wang},
  {Luo}, {Yang}, {Jing}, {Wang}, and {Mo}}}]{Xia:2017}
\bibinfo{author}{\bibfnamefont{Q.}~\bibnamefont{{Xia}}},
  \bibinfo{author}{\bibfnamefont{X.}~\bibnamefont{{Kang}}},
  \bibinfo{author}{\bibfnamefont{P.}~\bibnamefont{{Wang}}},
  \bibinfo{author}{\bibfnamefont{Y.}~\bibnamefont{{Luo}}},
  \bibinfo{author}{\bibfnamefont{X.}~\bibnamefont{{Yang}}},
  \bibinfo{author}{\bibfnamefont{Y.}~\bibnamefont{{Jing}}},
  \bibinfo{author}{\bibfnamefont{H.}~\bibnamefont{{Wang}}}, \bibnamefont{and}
  \bibinfo{author}{\bibfnamefont{H.}~\bibnamefont{{Mo}}},
  \bibinfo{journal}{\apj} \textbf{\bibinfo{volume}{848}}, \bibinfo{eid}{22}
  (\bibinfo{year}{2017})

\bibitem[{\citenamefont{{Okumura} et~al.}(2012)\citenamefont{{Okumura},
  {Seljak}, and {Desjacques}}}]{Okumura:2012b}
\bibinfo{author}{\bibfnamefont{T.}~\bibnamefont{{Okumura}}},
  \bibinfo{author}{\bibfnamefont{U.}~\bibnamefont{{Seljak}}}, \bibnamefont{and}
  \bibinfo{author}{\bibfnamefont{V.}~\bibnamefont{{Desjacques}}},
  \bibinfo{journal}{\jcap} \textbf{\bibinfo{volume}{11}}, \bibinfo{eid}{014}
  (\bibinfo{year}{2012})

\bibitem[{\citenamefont{{Schneider} and {Bridle}}(2010)}]{Schneider:2010}
\bibinfo{author}{\bibfnamefont{M.~D.} \bibnamefont{{Schneider}}}
  \bibnamefont{and} \bibinfo{author}{\bibfnamefont{S.}~\bibnamefont{{Bridle}}},
  \bibinfo{journal}{\mnras} \textbf{\bibinfo{volume}{402}},
  \bibinfo{pages}{2127} (\bibinfo{year}{2010})

\bibitem[{\citenamefont{{Navarro} et~al.}(1996)\citenamefont{{Navarro},
  {Frenk}, and {White}}}]{Navarro:1996}
\bibinfo{author}{\bibfnamefont{J.~F.} \bibnamefont{{Navarro}}},
  \bibinfo{author}{\bibfnamefont{C.~S.} \bibnamefont{{Frenk}}},
  \bibnamefont{and} \bibinfo{author}{\bibfnamefont{S.~D.~M.}
  \bibnamefont{{White}}}, \bibinfo{journal}{\apj}
  \textbf{\bibinfo{volume}{462}}, \bibinfo{pages}{563} (\bibinfo{year}{1996})

\bibitem[{\citenamefont{{Okumura} et~al.}(2015)\citenamefont{{Okumura}, {Hand},
  {Seljak}, {Vlah}, and {Desjacques}}}]{Okumura:2015}
\bibinfo{author}{\bibfnamefont{T.}~\bibnamefont{{Okumura}}},
  \bibinfo{author}{\bibfnamefont{N.}~\bibnamefont{{Hand}}},
  \bibinfo{author}{\bibfnamefont{U.}~\bibnamefont{{Seljak}}},
  \bibinfo{author}{\bibfnamefont{Z.}~\bibnamefont{{Vlah}}}, \bibnamefont{and}
  \bibinfo{author}{\bibfnamefont{V.}~\bibnamefont{{Desjacques}}},
  \bibinfo{journal}{\prd} \textbf{\bibinfo{volume}{92}}, \bibinfo{eid}{103516}
  (\bibinfo{year}{2015})

\bibitem[{\citenamefont{{Rasmussen} and {Williams}}(2006)}]{Rasmussen:2006}
\bibinfo{author}{\bibfnamefont{C.~E.} \bibnamefont{{Rasmussen}}}
  \bibnamefont{and} \bibinfo{author}{\bibfnamefont{C.~K.~I.}
  \bibnamefont{{Williams}}}, \emph{\bibinfo{title}{{Gaussian Processes for
  Machine Learning}}} (\bibinfo{year}{2006}).

\bibitem[{\citenamefont{{Ambikasaran} et~al.}(2015)\citenamefont{{Ambikasaran},
  {Foreman-Mackey}, {Greengard}, {Hogg}, and {O'Neil}}}]{Ambikasaran:2015}
\bibinfo{author}{\bibfnamefont{S.}~\bibnamefont{{Ambikasaran}}},
  \bibinfo{author}{\bibfnamefont{D.}~\bibnamefont{{Foreman-Mackey}}},
  \bibinfo{author}{\bibfnamefont{L.}~\bibnamefont{{Greengard}}},
  \bibinfo{author}{\bibfnamefont{D.~W.} \bibnamefont{{Hogg}}},
  \bibnamefont{and} \bibinfo{author}{\bibfnamefont{M.}~\bibnamefont{{O'Neil}}},
  \bibinfo{journal}{IEEE Transactions on Pattern Analysis and Machine
  Intelligence} \textbf{\bibinfo{volume}{38}} (\bibinfo{year}{2015})

\bibitem[{\citenamefont{{Lapi} and {Cavaliere}}(2009)}]{Lapi:2009}
\bibinfo{author}{\bibfnamefont{A.}~\bibnamefont{{Lapi}}} \bibnamefont{and}
  \bibinfo{author}{\bibfnamefont{A.}~\bibnamefont{{Cavaliere}}},
  \bibinfo{journal}{\apj} \textbf{\bibinfo{volume}{692}}, \bibinfo{pages}{174}
  (\bibinfo{year}{2009})

\bibitem[{\citenamefont{{Faltenbacher} and {White}}(2010)}]{Faltenbacher:2010}
\bibinfo{author}{\bibfnamefont{A.}~\bibnamefont{{Faltenbacher}}}
  \bibnamefont{and} \bibinfo{author}{\bibfnamefont{S.~D.~M.}
  \bibnamefont{{White}}}, \bibinfo{journal}{\apj}
  \textbf{\bibinfo{volume}{708}}, \bibinfo{pages}{469} (\bibinfo{year}{2010})

\bibitem[{\citenamefont{{Spergel} and {Steinhardt}}(2000)}]{Spergel:2000}
\bibinfo{author}{\bibfnamefont{D.~N.} \bibnamefont{{Spergel}}}
  \bibnamefont{and} \bibinfo{author}{\bibfnamefont{P.~J.}
  \bibnamefont{{Steinhardt}}}, \bibinfo{journal}{Physical Review Letters}
  \textbf{\bibinfo{volume}{84}}, \bibinfo{pages}{3760} (\bibinfo{year}{2000})

\bibitem[{\citenamefont{{Okumura, et al.}}(2018)}]{Okumura:2018}
\bibinfo{author}{\bibfnamefont{T.}~\bibnamefont{{Okumura, et al.}}},
  \bibinfo{journal}{in preparation}  (\bibinfo{year}{2018}).

\bibitem[{\citenamefont{{Okumura}
  et~al.}(2017{\natexlab{b}})\citenamefont{{Okumura}, {Nishimichi}, {Umetsu},
  and {Osato}}}]{Okumura:2017a}
\bibinfo{author}{\bibfnamefont{T.}~\bibnamefont{{Okumura}}},
  \bibinfo{author}{\bibfnamefont{T.}~\bibnamefont{{Nishimichi}}},
  \bibinfo{author}{\bibfnamefont{K.}~\bibnamefont{{Umetsu}}}, \bibnamefont{and}
  \bibinfo{author}{\bibfnamefont{K.}~\bibnamefont{{Osato}}},
  \bibinfo{journal}{ArXiv e-prints}  (\bibinfo{year}{2017}{\natexlab{b}}),
  \eprint{1706.08860}.

\end{thebibliography}
 \end{document}